# Pressure-induced Superconductivity in Tin Sulfide

*Ryo Matsumoto[a,b], Peng Song[a,b], Shintaro Adachi[a], Yoshito Saito[a,b], Hiroshi Hara[a,b], Kazuki Nakamura[c], Sayaka Yamamoto[a,b,c], Hiromi Tanaka[c], Tetsuo Irifune[d], Hiroyuki Takeya[a], and Yoshihiko Takano[a,b]*

[a]*International Center for Materials Nanoarchitectonics (MANA), National Institute for Materials Science, 1-2-1 Sengen, Tsukuba, Ibaraki 305-0047, Japan*
[b]*University of Tsukuba, 1-1-1 Tennodai, Tsukuba, Ibaraki 305-8577, Japan*
[c]*National Institute of Technology, Yonago College, 4448 Hikona, Yonago, Tottori 683-8502, Japan*
[d]*Geodynamics Research Center, Ehime University, Matsuyama, Ehime 790-8577, Japan*

**Abstract**

Tin sulfide (SnS) was successfully synthesized in single crystals by a melt and slow-cooling method. The obtained sample was characterized by an X-ray diffraction, an energy dispersive spectrometry, and an X-ray photoelectron spectroscopy. Electrical transport properties in SnS were investigated under high pressure using a diamond anvil cell with boron-doped metallic diamond electrodes and undoped diamond insulating layer. We successfully observed an insulator to metal transition from 12.5 GPa and pressure-induced superconductivity at 5.8 K under 47.8 GPa as predicted by a theoretical calculation.

Tin chalcogenide binary compounds have been studied as superior functional materials with a high thermoelectric performance in the past several years [1-7]. Especially, tin selenide (SnSe) shows remarkable thermoelectric property of an ultrahigh figure of merit $ZT$ value of 2.6 at 923 K, due to its low thermal conductivity, controllable resistivity, and high Seebeck coefficient [1,8]. Also, superconductivity in tin selenide was recently reported under high pressure [9]. Tin telluride similarly shows superior thermoelectric property as lead-free materials [10]. The carrier-doped SnTe shows superconductivity around 2 K under ambient pressure [11]. These indicate that tin chalcogenides binary compounds are new vein of superconducting materials.

According to recent first principles calculations, superconductivity in tin sulfide binary compounds SnS and $Sn_3S_4$ are predicted under high pressure [12]. A stable phase at ambient pressure α-SnS-*Pnma* with a band gap of 1.28 eV first changes to metallic β-SnS-*Cmcm* phase from 9 GPa. Under further compression, a superconducting γ-SnS-*Pm-3m* phase appears from 40 GPa with transition temperatures ($T_c$) of 9.74 K. A higher $T_c$ compound of $Sn_3S_4$ with *I-43d* structure shows instability at ambient pressure. The metallic phase is stable from 15 GPa and maximum $T_c$ of 21.9 K appears from 30 GPa. Although similar structural transition and superconductivity were experimentally observed in related materials SnSe [9] and $Sn_3Se_4$ [13], there is no report regarding to tin sulfide.

In this study, we experimentally confirm the predicted superconductivity in tin sulfide. Single crystals of tin sulfide were synthesized via conventional melt and slow-cooling method. Crystal structures of obtained samples were analyzed by an X-ray diffraction (XRD) using a Mini Flex 600 (Rigaku) with Cu Kα radiation. The compositional ratios were investigated by an energy dispersive spectrometry (EDX) using a JSM-6010LA (JEOL). An X-ray photoelectron spectroscopy (XPS) analyses using an AXIS-ULTRA DLD (Shimadzu/Kratos) with Al Kα X-ray radiation ($hv$ = 1486.6 eV) were carried out to clear surface states of the obtained crystals. The XPS measurements were operated under a pressure of the order of $10^{-9}$ Torr. Photoelectron peaks were analyzed by pseudo-Voigt functions peak fitting with a background subtraction by an active Shirley method using a COMPRO software [14]. Superconductivity was examined by electrical transport measurements via a standard four probe method under high pressure using an originally designed diamond anvil cell [15,16]. Cubic boron nitride powders with a ruby manometer were used as a pressure-transmitting medium. Applied pressures were estimated by a fluorescence from ruby powders [17] and a Raman spectrum from a culet of the top diamond anvil [18] by an inVia Raman Microscope (RENISHAW).

We tried to synthesize SnS and $Sn_3S_4$ in single crystals by a melt and slow-cooling method. Starting materials of Sn grains (99.99%) and S grains (99.99%) were put into evacuated quartz tubes in stoichiometric compositions of SnS and $Sn_3S_4$. The ampoules were heated at 350ºC for 4 hours, subsequently at 900ºC for 20 hours, and slowly cooled to 880ºC for 20 hours followed by a furnace cooling.

Figure 1(a) shows an XRD pattern of well-ground SnS crystals. All observed peaks were well indexed to an orthorhombic *Pnma* structure. Figure 1(b) shows an XRD pattern of one piece of the obtained SnS crystal. The pattern only exhibits *h*00 diffraction peaks indicating that the sample is high quality single crystal. The EDX analysis showed a composition of Sn 52.04% and S 47.96%.

The Sn-rich quasi-stoichiometric composition in SnS is consistent with previous report in some literatures [19,20]. On the other hand, the obtained crystals from the nominal composition of $Sn_3S_4$ contained SnS, $SnS_2$ and $Sn_2S_3$ components without the desired material of $Sn_3S_4$. To obtain the $Sn_3S_4$ phase that would show higher $T_c$ of 21.9 K by the theoretical prediction, a high pressure synthesis above 15 GPa should be required [12].

Electrical resistivity measurements for the obtained SnS single crystal were carried out at ambient pressure via a standard four prove method. Figure 1(c) shows a temperature dependence of a resistivity for SnS. The order of resistivity of 50 Ω·cm at 300 K is consistent with a typical value of SnS [21]. To evaluate an activation energy of the obtained SnS, the measured resistivity is fitted by the Arrhenius relationship of $\rho=\rho_0 \times \exp(E_a/k_B T)$, where $\rho_0$ is a constant residual resistivity value, $E_a$ is the activation energy, $k_B$ is a Boltzmann constant, and $T$ is the temperature, as shown in the inset of Fig. 1(c). The straight-line fit yields the $E_a$ value near room temperature of 0.12 eV, which shallow level shows an agreement with a tendency in the Sn-rich SnS [22,23]. Figure 1(d) shows a Hall voltage of SnS as a function of applied magnetic field to comfirm a carrier type and a carrier concentration at room temperature and ambient pressure. From the slope of Hall voltage versus magnetic field, the carrier concentrations of samples have been calculated using the formula, $(V_H/I) = (1/ned)H$, where $V_H$ is the Hall voltage, $I$ is a current, $n$ is a number of carriers, $e$ is an elementary charge, $H$ is the magnetic field and $d$ is a sample thickness. The $V_H$ curve showed a positive slope, indicating a p-type characteristic with the carrier concentration of $6.5 \times 10^{17}$ cm$^{-3}$. Although a typical carrier concentration in SnS is the order of $10^{15}$ cm$^{-3}$ [24], it could be increased up to the order of $10^{17}$ cm$^{-3}$ by a quasi-stoichiometric composition [19], which is consistent with our EDX result. The relatively high carrier concentration would contribute to the observed smaller energy gap in the synthesized SnS. Also, this carrier concentration is slightly higher than that of a hole-doped SnSe of $4.5 \times 10^{17}$ cm$^{-3}$ [25].

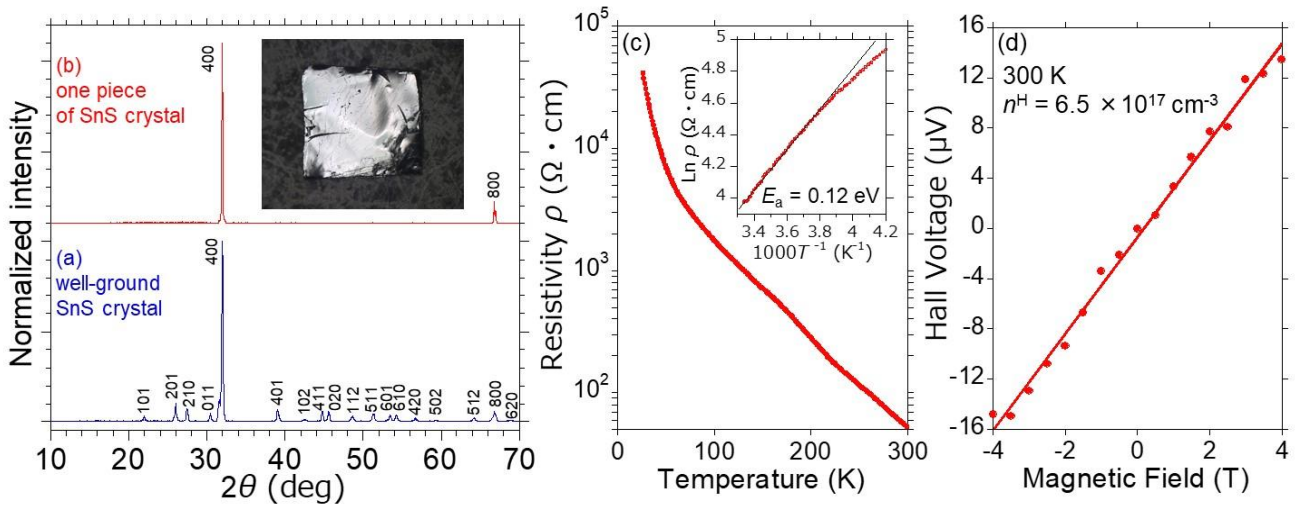

**Figure 1. XRD patterns of (a) well-ground SnS and (b) one piece of SnS single crystal, (c) Temperature dependence of resistivity for SnS at ambient pressure. The inset shows Arrhenius plot for resistivity near room temperature, (c) Hall voltage of SnS as a function of applied magnetic field.**

High-resolution XPS measurements of Sn 3*d* orbital were carried out to investigate a surface state of the obtained SnS single crystal. Figure 2(a) shows a Sn 3*d* spectrum of a SnS single crystal without cleaving treatment. The spectrum was deconvoluted into four peaks as labeled in the fig. 2(a). The main components of peak 1 at 485.7 eV and peak 2 (centered at 494.1 eV) are corresponding to a $Sn^{2+}$ valence state with a spin–orbit splitting of 8.4 eV [26]. The shoulder components of peak3 (centered at 486.9 eV) and peak4 (centered at 495.3 eV) are originated from a $Sn^{4+}$ oxidization state because a chemical shift of 0.7-1.5 eV between $Sn^{4+}$ and $Sn^{2+}$ was accordingly reported by previous studies [27,28].

Figure 2(b) shows the spectrum from a cleaved SnS single crystal using scotch tape in a highly vacuumed pre-chamber of the order of $10^{-7}$ Torr to obtain its intrinsic valence state. In both Sn $3d_{5/2}$ and Sn $3d_{3/2}$ orbitals, only sharp peaks were observed at 485.7 eV and 494.1 eV according to the $Sn^{2+}$ state. These results indicate that an outermost surface of the SnS is oxidized by the air, which is simular to the related material SnSe [25]. Here, a thickness of the $Sn^{4+}$ layer was estimated using an equation, $d=L\cos\theta\ln(I_A/I_B+1)$ [14], where *d* is the thickness, *L* is an inelastic mean free path of a photoelectron from the sample, $\theta$ is an emission angle, $I_A$ and $I_B$ are peak area intensities from $Sn^{4+}$ and $Sn^{2+}$ peaks, respectively. The value of *L* is 2.4 nm [29], $\theta$ is 0°, and $I_A/I_B$ is 0.63. The estimation indicates that only 1.2 nm depth from the surface is the valence state of $Sn^{4+}$, and the valence state of the bulk is $Sn^{2+}$. We can easily evaluate an intrinsic property of SnS by cleaving the sample surface before transport measurements.

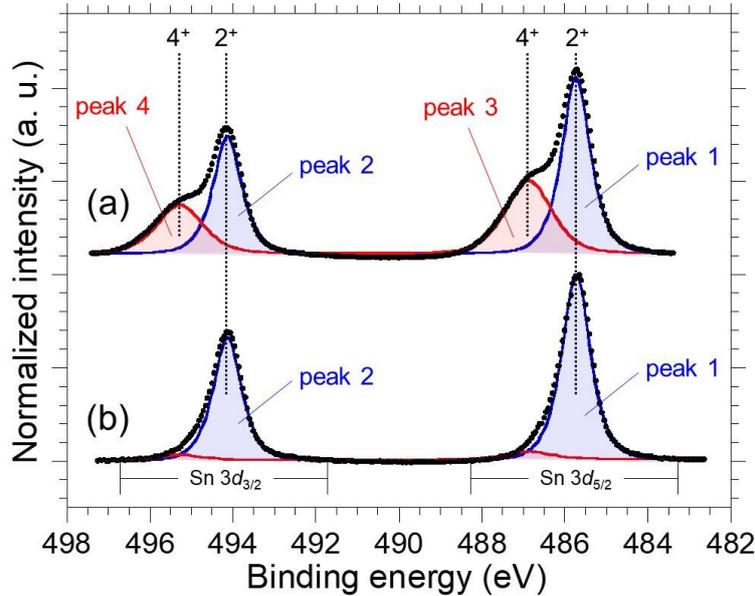

**Figure 2. High-resolution XPS measurements of Sn 3d orbital of (a) SnS single crystal without cleaving treatment, and (b) cleaved SnS single crystal in a highly vacuumed pre-chamber of the order of $10^{-7}$ Torr using scotch tape.**

To examine the predicted superconductivity in SnS, a temperature dependence of resistance was measured under high pressures. Here, we used the originally designed diamond anvil cell [15,16] in the high pressure experiments as shown in a schematic image of Fig. 3(a). Sample voltage is detected by the heavily boron-doped metallic diamond (BDD) electrodes on the bottom anvil. The electrodes and the metal gasket are electrically separated by the insulating undoped diamond (UDD) layer. The details of a fabrication process of these special diamonds are described in the literatures [15,16]. In the measurement of SnS, the 6 prove design of electrodes was used as shown in Fig. 3(b). Figure 3(c) is an optical image of the one piece of single crystal that was placed on the center of the bottom diamond anvil with the boron-doped diamond electrodes. The crystal was cleaved by the scotch tape just before a pressure application to remove the oxidized surface.

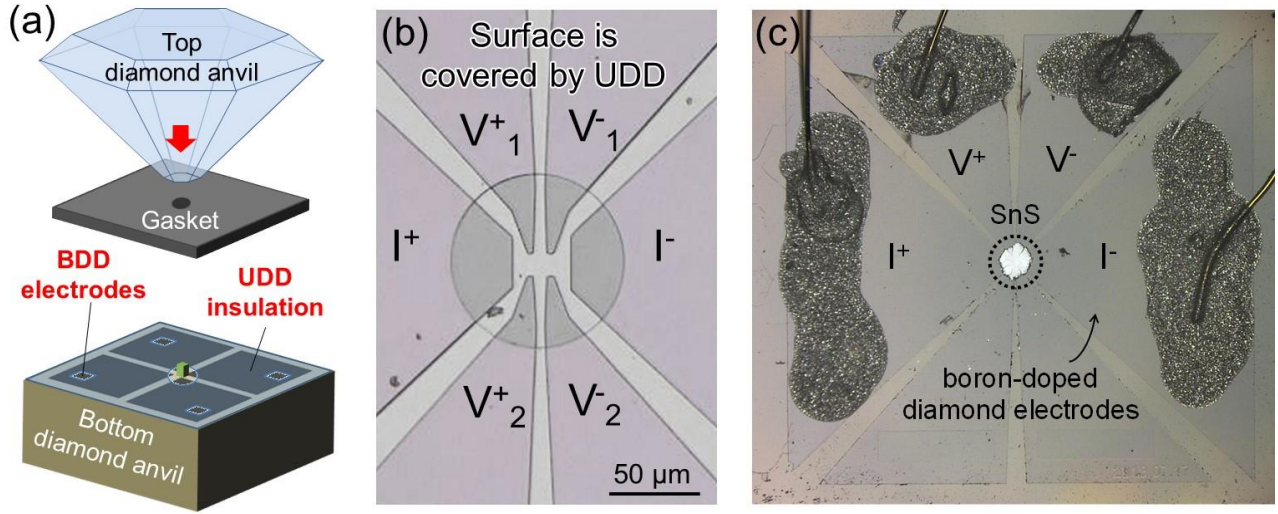

**Figure 3. (a) Originally designed diamond anvil cell with boron-doped diamond (BDD) electrodes and undoped diamond (UDD) insulating layer. (b) Enlargement of 6 prove electrodes, (c) Sample space of bottom diamond anvil with one piece of SnS single crystal.**

Figure 4 (a) shows the temperature dependences of resistance in SnS under various pressures up to 47.8 GPa. The sample first showed insulating behavior under 4 GPa. The resistance was drastically decreased with an increase of the applied pressure up to 12.5 GPa. Figure 4 (b) exhibits an estimated energy gap from the Arrhenius plots for resistance of $R=R_0\times\exp(E_a/k_BT)$ as a function of the applied pressure. The Arrhenius plots are inserted in the figure. The energy gap was decreased with the increase of pressure with a slope of 18 meV/GPa, and almost closed under 12.5 GPa. When the pressure achieved to 23.6 GPa, we observed a negative slope of $dR/dT$, namely pressure-induced insulator to metal transition. Under further pressure region at 47.8 GPa, the sample exhibited a sudden drop of resistance from 5.8 K corresponding to superconductivity. The pressure-induced insulator to metal transition and the superconductivity were successfully observed based on the theoretical prediction. Although it can be expected an increase of $T_c$ under higher pressure range because the superconductivity is still filamentary, the diamond anvil was broken after the measurement, unfortunately.

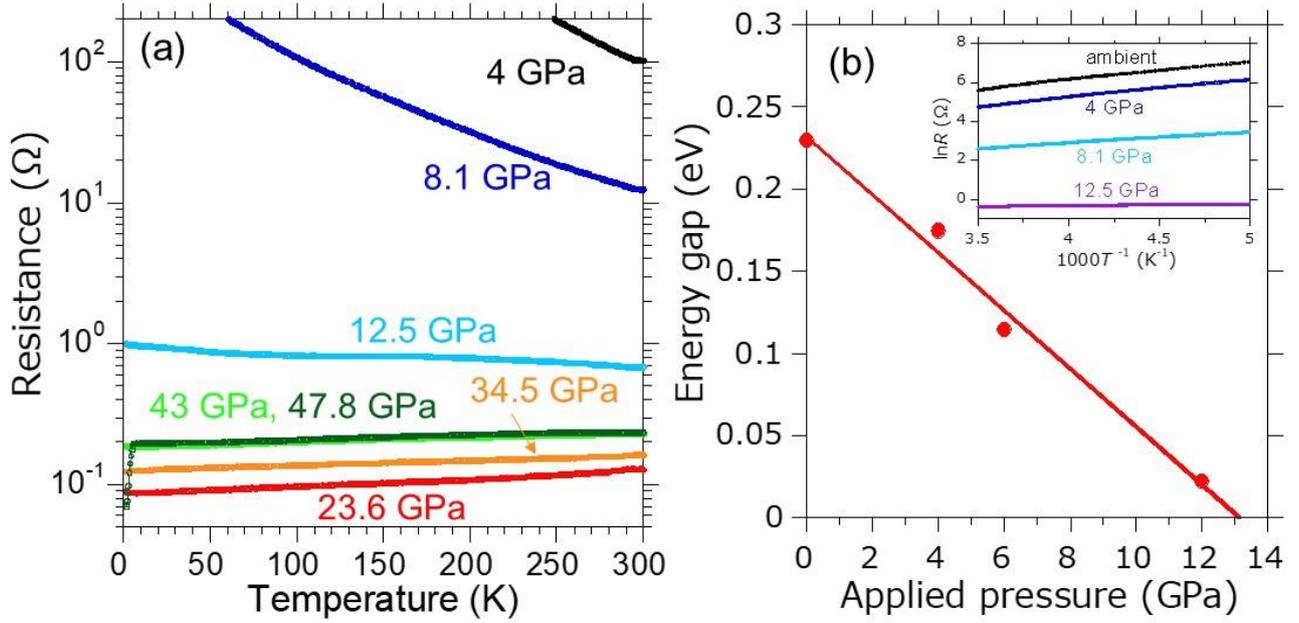

**Figure 4.** (a) Temperature dependence of resistance in SnS single crystal under various pressures up to 47.8 GPa, (b) Estimated energy gap of SnS as a function of applied pressure. The inset shows Arrhenius plots for resistance under 4 GPa, 8.1 GPa, and 12.5 GPa.

To further confirm that the drop of resistance in SnS under 47.8 GPa is originated from the superconductivity, we measured the temperature dependence of resistance under magnetic field up to 1.5 T with a direction perpendicular to a [100] plane, as shown in Fig. 5(a). The drop of resistance was gradually suppressed by an increase of the applied magnetic field, and completely disappeared above 2 K under 1.5 T. This suppression indicates the drop of resistance in SnS under 47.8 GPa comes from the superconductivity. Figure 5(b) shows a temperature dependence of upper critical field $H_{c2}$ estimated from the Werthamer-Helfand-Hohenberg (WHH) approximation [30] for Type II superconductors in a dirty limit. The extrapolated $H_{c2}(0)$ were 1.6 T under 47.8 GPa. From the Ginzburg–Landau (GL) formula $H_{c2}(0) = \Phi_0/2\pi\xi(0)^2$, where the $\Phi_0$ is a fluxoid, the $\xi(0)$ is a coherence length at zero temperature, $\xi(0)$ is 1.4 nm. According to the above investigation, SnS exhibits higher $T_c$ and $H_{c2}(0)$ of 5.8 K and 1.6 T than those of 3.2 K and 1.1 T in superconducting SnSe with *Pm-3m* structure [9]. Assuming a phonon-mediated Bardeen-Cooper-Schrieffer (BCS) superconductivity in γ-SnS-*Pm-3m* structure above 40 GPa as pointed in the literature [12], a high density of state (DOS) near the Fermi level and lighter atomic mass possibly contribute the higher $T_c$. Since hole-doped SnSe shows higher $T_c$ and $H_{c2}(0)$ compared with those of undoped SnSe [25], it can be expected similar enhancement of superconducting properties in carrier doped SnS.

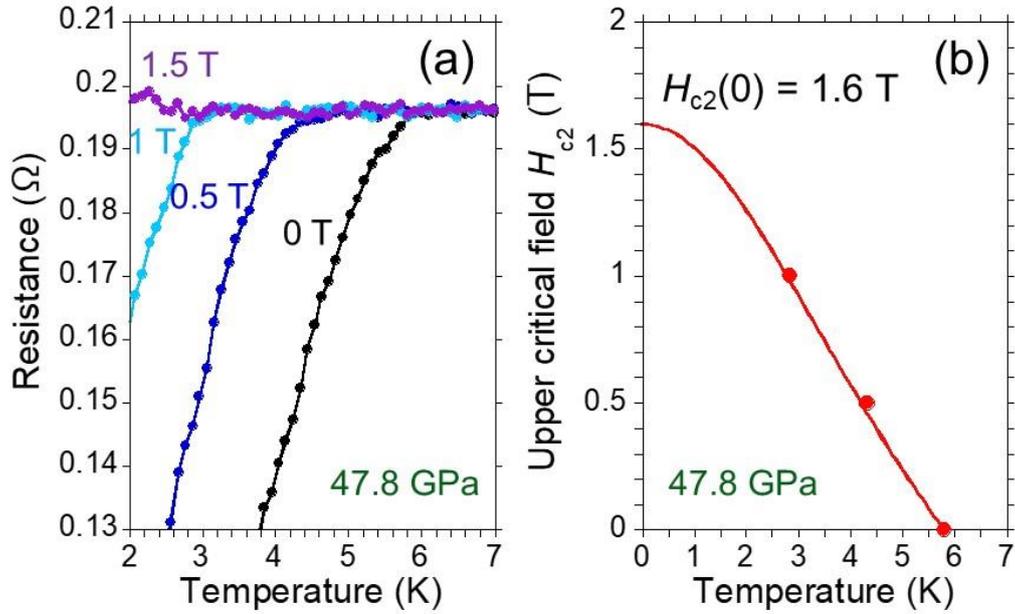

**Figure 4.** (a) Temperature dependence of resistance at 47.8 GPa under magnetic field up to 1.5 T with a direction perpendicular to a [100] plane. (b) Temperature dependence of upper critical field $H_{c2}$ estimated from the Werthamer-Helfand-Hohenberg (WHH) approximation.

During this study, we succeeded in synthesis of SnS single crystal. The resistance measurement under high pressure using our originally designed diamond anvil cell revealed the insulator to metal transition from 12.5 GPa, and superconductivity at 5.8 K under 50 GPa with correspondence to the theoretical prediction. Such a theory-preceding exploration for superconductors, for example, data-driven approaches [31,32], discovery of hydrogen-rich high-$T_c$ superconductors [33-36], and so on, would be more accelerated in further materials science.

**Acknowledgment**

This work was partly supported by JST CREST Grant No. JPMJCR16Q6, JST-Mirai Program Grant Number JPMJMI17A2, and JSPS KAKENHI Grant Number JP17J05926. A part of the fabrication process was supported by NIMS Nanofabrication Platform in Nanotechnology Platform Project sponsored by the Ministry of Education, Culture, Sports, Science and Technology (MEXT), Japan. The part of the high pressure experiments were supported by the Visiting Researcher's Program of Geodynamics Research Center, Ehime University.